%% file: main.tex
\documentclass[USenglish,oneside,twocolumn]{article}
\usepackage[utf8]{inputenc}
\usepackage[big]{dgruyter_NEW}
\usepackage[hyphens]{url}
\usepackage{pifont}
\usepackage{comment}
\newcommand{\checkmark}{\ding{51}}
\newcommand{\xmark}{\ding{55}}
\usepackage{multirow}
\usepackage{multicol}
\usepackage{natbib}
\usepackage{enumitem}
\usepackage{fontawesome}
\usepackage{mathabx}
\include{localdefs}

\usepackage{adjustbox}
\usepackage{array}

\newcolumntype{R}[2]{%
    >{\adjustbox{angle=#1,lap=\width-(#2)}\bgroup}%
    l%
    <{\egroup}%
}
\newcommand*\rot{\multicolumn{1}{R{60}{1em}}}

\usepackage{tikz}
\newcommand*\emptycirc[1][1ex]{%
    \begin{tikzpicture}[baseline=-\the\dimexpr\fontdimen22\textfont2\relax]
    \draw (0,0) circle (#1); 
    \end{tikzpicture}}
\newcommand*\dotcirc[1][1ex]{%
    \begin{tikzpicture}[baseline=-\the\dimexpr\fontdimen22\textfont2\relax]
    \draw[fill] (0,0) circle (#1/3) ;
    \draw (0,0) circle (#1);
    \end{tikzpicture}}
\newcommand*\halfcirc[1][1ex]{%
    \begin{tikzpicture}[baseline=-\the\dimexpr\fontdimen22\textfont2\relax]
    \draw[fill] (0,0)-- (90:#1) arc (90:270:#1) -- cycle ;
    \draw (0,0) circle (#1);
    \end{tikzpicture}}
\newcommand*\fullcirc[1][1ex]{%
    \begin{tikzpicture}[baseline=-\the\dimexpr\fontdimen22\textfont2\relax]
    \draw[fill] (0,0) circle (#1); 
    \end{tikzpicture}} 
    
\newcommand\Ua{$\Uparrow$}
\newcommand\ua{$\uparrow$}
\newcommand\da{$\downarrow$}
\newcommand\Da{$\Downarrow$}
\newcommand\Uda{$\Updownarrow$}
\newcommand\uda{$\updownarrow$}

\journalname{Proceedings on Privacy Enhancing Technologies}
\startpage{1}
\journalyear{2022}
\journalvolume{..}
\journalissue{..}

\cclogo{\includegraphics{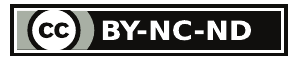}}

\newcommand{\ThreatModelName}{Post-Compromise Data Confidentiality}
\newcommand{\Threatmodelname}{Post-compromise data confidentiality}
\newcommand{\threatmodelname}{post-compromise data confidentiality}

\newcommand{\defenseparadigm}{trusted hardware-anchored cloud services}

\begin{document}

\author*[1]{Maximilian Zinkus}
\author[2]{Tushar M. Jois}
\author[2]{Matthew Green}
\affil[1]{Johns Hopkins University, zinkus@cs.jhu.edu}
\affil[2]{Johns Hopkins University, \{jois, mgreen\}@cs.jhu.edu}

\title{\huge SoK: Cryptographic Confidentiality of Data on Mobile Devices}
\runningtitle{SoK: Cryptographic Confidentiality of Data on Mobile Devices}

\begin{abstract}
{Mobile devices have become an indispensable component of modern life. Their high storage capacity gives these devices the capability to store vast amounts of sensitive personal data, which makes them a high-value target: these devices are routinely stolen by criminals for data theft, and are increasingly viewed by law enforcement agencies as a valuable source of forensic data. Over the past several years, providers have deployed a number of advanced cryptographic features intended to protect data on mobile devices, even in the strong setting where an attacker has physical access to a device. Many of these techniques draw from the research literature, but have been adapted to this entirely new problem setting.\\
This involves a number of novel challenges, which are incompletely addressed in the literature. In this work, we outline those challenges, and systematize the known approaches to securing user data against extraction attacks. Our work proposes a methodology that researchers can use to analyze cryptographic data confidentiality for mobile devices. We evaluate the existing literature for securing devices against data extraction adversaries with powerful capabilities including access to devices and to the cloud services they rely on. We then analyze existing mobile device confidentiality measures to identify research areas that have not received proper attention from the community and represent opportunities for future research.}
\end{abstract}


  \journalname{Proceedings on Privacy Enhancing Technologies}
 \DOI{Editor to enter DOI}
%
 
\maketitle
\section{Introduction}
Mobile devices have become an indispensable component of modern life. Projections estimate that there are as many as four billion smartphone users~\cite{statista_number}. Simultaneously, smartphone connectivity, data storage, and sensing capabilities continue to improve. The adoption of these devices has complex implications for user data privacy: the smartphone vastly increases the amount of personal information that individuals can carry on their person, while simultaneously exposing that data to an unprecedented risk of theft. Portability and ease of physical access makes smartphones a target for malicious actors and law enforcement alike: to the former it provides new opportunities for criminality~\cite{icloudleaks,us_cert_thefts,dhsreportphones}, and to the latter it offers new avenues for investigation and surveillance~\cite{elcomsoft_methods,comey14,pegasus_wapo,un_pegasus}. Further,
``the cloud'' has effectively become an extension of the device itself~\cite{matzner2014privacy}, and can enable data extraction even when physical devices are kept physically secure~\cite{pi_cloud_extraction,elcomsoft_tokens,cellebrite_cloud}.

\begin{table*}[t]
    \caption{Examples of post-compromise data confidentiality attacks and mitigations. Each mitigation defends against equivalent classes of \emph{illicit} and \emph{lawful} access, unable to distinguish them (\S\ref{sec:background}).}
    \centering
    \begin{tabular}{p{0.3\linewidth} c p{0.3\linewidth} | p{0.3\linewidth}}
         Illicit Access Attack & & Lawful Access Attack & Protections \\
         \hline \hline
         Logical extraction via jailbreak/zero-day  & \(\sim\) & Logical extraction via UFED or GrayKey & FDE, RTFE (\S\ref{sec:storage_enc}) \\
         \hline
         Cloud data retrieval via provider compromise or insider threat & \(\sim\) & Cloud data extraction via court order & User-Controlled Encryption (\S\ref{sec:cloud_uce}) \\
         \hline
         Manual extraction via user coercion or blackmail & \(\sim\) & Manual extraction via user interrogation & Deniable Encryption (\S\ref{sec:storage_enc}), Trusted Hardware (\S\ref{sec:device_auth}, \S\ref{sec:cloud_utca}) \\
         \hline
         Physical extraction from stolen device & \(\sim\) & Physical extraction from device in evidence & FDE, RTFE, Secure Deletion, Deniable Encryption (\S\ref{sec:storage_enc}) \\
         \hline
         On-path protection downgrade by MitM & \(\sim\) & Selective protection downgrade from compelled provider & Transparency (\S\ref{sec:transparency}) \\
         \hline
         Brute-force password guessing & $=$ & Brute-force password guessing & Trusted Hardware (\S\ref{sec:device_auth}, \S\ref{sec:cloud_utca}) \\
    \end{tabular}
    \label{tab:lawful_illicit}
\end{table*}

\paragraph{Scope of this work} The space of ``mobile device security and privacy'' is vast, and has been extensively studied in the research literature. In practice, this area includes a large set of technology areas, including software security, hardware security, access control, and network security. Further, privacy \textit{against device software} is commonly considered~\cite{spensky2016sok}. An area that has received less study, however, is a specific layer of the mobile device security stack: the {\em cryptographic systems that protect data confidentiality}. These user-controlled cryptographic systems have increasingly been included into mobile devices as a form of ``defense in depth'' to provide robustness when other layers fail; we refer to this as \textit{\threatmodelname}.

The inclusion of these cryptographic systems reflects a growing understanding in industry that traditional software-based access control has proven insufficient to stop real-world attacks --- an understanding that is strongly backed by the historical record~\cite{full_paper}. The importance of these systems has only increased as data shifts to cloud-based servers, which are vulnerable to remote attacks. Put succinctly: the focus of this work is the protection of user data stored in a locked (i.e. passcode-protected) modern mobile device, against adversaries who may physically seize the device and may target associated cloud services. We consider illicit attackers, but must also include legally compelled or forensic access -- government and law enforcement access has deeply influenced mobile device and cloud service development, and from a technical perspective the threats to user data confidentiality are indistinguishable (\S\ref{sec:background}).

\paragraph{Purpose of this work} While encryption technology historically emerged from the research community to industry for refinement, this trend has reversed in recent years. Deployment of mobile device encryption has become a priority for manufacturers and vendors~\cite{full_paper}. As such, many prevalent techniques and threat models have been developed outside of the research community, and may not have been clearly documented or analyzed by researchers. In this work we attempt to rectify the problem: our goal is to capture and systematize the latest developments in this dynamic area, and to provide the research community with an overview of the current open research areas.

\paragraph{Contributions} In this work we systematize the current state of the art in cryptographic protection for mobile devices and their associated cloud services. Our goal is to integrate the latest developments from industrial security systems with knowledge developed by the research community, to assist the community in identifying the open research problems that exist in this area. Towards this goal we contribute the following:
\begin{itemize}
    \setlength\itemsep{0.25em}
    \item[\ding{99}] We thoroughly examine the research literature in defending mobile device data against data extraction attacks.
    \item[\ding{99}] We contextualize our work through extensive analysis of industry literature, as well as government documents and standards, technical documents, articles, and other sources.
    \item[\ding{99}] We formalize a novel {threat model} for mobile devices considering the subtleties of cloud integration, and an emerging {defense paradigm} for cloud services.
    \item[\ding{99}] We systematize \textit{cryptographic confidentiality mechanisms} from the research literature and from industry solutions.
    \item[\ding{99}] We identify remaining technical challenges towards achieving mobile device and cloud data privacy against data extraction adversaries, providing motivation and direction for future work grounded in the realities of the modern device ecosystem.
\end{itemize}

\section{Background}\label{sec:background}

\paragraph{Post-compromise security} In this work, we analyze confidentiality mechanisms which mitigate data extraction attacks. These attacks occur in the setting where devices and cloud services may be compromised via software, hardware, or procedural vulnerabilities. Thus, we consider the extent of confidentiality which remains \textit{given} such compromise. Post-compromise cryptography has been considered for data in-transit~\cite{e2e_sok}; we consider an analogous notion for data at-rest on cloud-enabled mobile devices.

\paragraph{The central role of encryption} Encryption is integral to post-compromise data confidentiality. Encryption systems rely on significantly smaller ``trusted code bases'' (TCBs) than other security systems, and are generally designed transparently as public standards. As a result, common vulnerabilities are often not present in encryption systems, or do not directly expose encrypted data. The post-compromise setting can be viewed as a strengthening of Kerckhoff's principle~\cite{shannon1949communication}: confidentiality remains when the enemy not only understands, but \textit{controls} the system, except for the secret key.

\paragraph{Impact on law enforcement} While provider improvements continue to enhance data confidentiality on mobile devices, they have provoked a backlash from the law enforcement community. This reaction is best exemplified by the FBI's ``Going Dark'' initiative~\cite{comey14} which seeks to increase law enforcement access to encrypted data via legislative and policy initiatives. These concerns have also motivated agencies to invest in acquiring technical means for bypassing smartphone security. This dynamic broke into the public consciousness during the 2016 ``Apple v. FBI'' controversy~\cite{ApplevFBI_apple,apple_letter_2016,ApplevFBI_fbi}, in which Apple contested an FBI demand 
to bypass technical security measures. A vigorous debate over these issues continues to this day~\cite{carnegie_report,upturn_mass_extraction,pi_deepdive}. Since 2015, in the US alone hundreds of thousands of forensic searches of mobile devices have been executed by over 2,000 law enforcement agencies across the country~\cite{upturn_mass_extraction}. These agencies use inexpensive industry tools implementing exploits to extract data~\cite{upturn_mass_extraction,vice_db_news,pi_deepdive}. This tug-of-war between device manufacturers, law enforcement, and forensics vendors has an important consequence for users: at any given moment, it is difficult to know which smartphone security features are operating as intended, and which can be bypassed.

\paragraph{Illicit vs. lawful access}
In this work we are focused on threats to user data confidentiality. In the broadest sense, this requires us to consider any adversary that wishes to access user content {\em without the explicit consent of the user}. Inevitably, such adversaries include both criminal attackers as well as authorized law enforcement agencies who operate with judicial oversight, as well as some agencies that operate in the ``gray area'' between the two. Numerous works have explored this dynamic~\cite{goingbright,abelson2015keys,savage2018lawful}; in the technical sections of this work we will focus primarily on the techniques used by an attacker, rather than considering the attacker's intent. Nonetheless it is important to capture the distinction.

\textit{Illicit access} refers to access by a criminal attacker, while \textit{lawful access} refers to the legally-sanctioned extraction of data from a device. Crucially, current data protection mechanisms in the literature cannot distinguish between illicit and lawful access. Table~\ref{tab:lawful_illicit} illustrates this relationship. Device providers use cryptography to attempt to prevent anyone other than the user from accessing the device -- inadvertently (or otherwise), these protections apply against both illicit and lawful actors. Several proposals have been advanced to bridge this divide by designing mechanisms that provide plaintext only to authorized parties~\cite{ozzie2019providing,savage2018lawful,wright2018crypto}. While these ``exceptional access'' techniques may enable lawful access, they have been criticized as potentially reducing the security of the underlying encryption mechanism~\cite{abelson2015keys,clear_bad}.
Solving this problem in the general case remains an open area of research~\cite{arleas}. As a result, when purely considering the technical means of extraction, we must consider both types of access.

\subsection{Understanding Data Confidentiality}\label{sec:data}
We consider three classes of stakeholders in mobile device data confidentiality: users; providers who manufacture and design devices and services, but also may be targets of attacks or be required to service subpoenas or searches; and adversaries, whose capabilities imply privacy requirements for data confidentiality. Table~\ref{tab:data} provides a summary listing the types of data commonly targeted for attack by adversaries and considered sensitive by users and providers.

\begin{table}[t]
    \centering
    \caption{Targeted Data in Device and Cloud Attacks}\label{tab:data}
    \begin{tabular}{l}
        \textbf{Data Type} \\\hline
        $\bullet$ Complete copies of device contents, if available \\
        Otherwise:\\
        $\bullet$ Communications (messages, emails) \\
        $\bullet$ Metadata (IP addresses, call logs) \\
        $\bullet$ Location data (GPS, IP-based) \\
        $\bullet$ Contacts/address books \\
        $\bullet$ Social media data (messages, posts) \\
        $\bullet$ Media (images, videos, audio) \\
    \end{tabular}
\end{table}

\paragraph{Adversaries} Due to the breadth and frequency of law enforcement access to mobile devices, extensive evidence exists of \textit{what data} is targeted by law enforcement. On the other hand, illicit activity is far more opaque. As such, we can use law enforcement access as a proxy for the sensitivity of data across both types of access. From a wide array of sources~\cite{full_paper,pi_deepdive,pi_cloud_extraction,upturn_mass_extraction,vice_db,apple_fbi_assist2020,nist_forensics_spec,dhs_forensics,earn_it,ios_significant_locations_scooped} we ascertain law enforcement prioritization of data for investigation and surveillance, detailed in Table~\ref{tab:data}.

\paragraph{Providers} Device and cloud service providers (collectively, ``providers'') such as Apple and Google make privacy decisions on behalf of their billions of consumers. Notably, a cloud and device provider are often the same entity. These privacy decisions include providing user-controlled (``end-to-end'') encryption~\cite{apple_platform_security}, file encryption~\cite{apple_platform_security,aosp_fulldisk,aosp_filebased}, and encrypted cloud storage and backup~\cite{apple_platform_security,kensinger2018google}. However, these providers have been illicitly accessed~\cite{icloudleaks} and also routinely acquiesce to legal requests for user data~\cite{apple_transparency,google_transp_2019}, and have even allegedly scuttled plans to add provider-inaccessible user-controlled encryption based on law enforcement backlash~\cite{mennicloud2020}. There is risk in centralizing such control over privacy decisions into two large companies: providers become a single point of storage for a trove of user data. Smaller, privacy-focused providers exist (e.g. Purism~\cite{purism}), but have failed to capture markets on the scale of Apple and Google.

\paragraph{Adversary-provider interaction} Three factors combine to create a unique interaction of adversaries and providers: $1)$ the privileged position of providers as handlers of sensitive data, $2)$ attacks which access provider servers, and $3)$ the economic incentives for providers to be perceived as trustworthy. As a result, adversaries and providers together can represent a form of honest-but-curious~\cite{goldreich2009foundations} (or ``semi-honest'') party, passively observing user data. We note this pattern commonly in practice, where providers turn over or unintentionally leak data~\cite{full_paper,apple_legal,pi_cloud_extraction}. In rarer cases, they can represent covert~\cite{aumann2007security} or even fully-malicious~\cite{goldreich2009foundations} adversaries willing to undertake deliberate measures, while potentially attempting to evade detection or blame.

\paragraph{An emerging defense paradigm} In response to this reality, providers including Apple~\cite{apple_icloud_security}, Google~\cite{google_titanm}, and even apps such as Signal~\cite{signal_svr} have turned to trusted hardware in the cloud to perform sensitive actions. If correctly configured and deployed, trusted hardware can protect data from even the provider themselves to mitigate these threats. We refer to services adopting this model as \textit{\defenseparadigm}, and discuss the use of trusted hardware extensively throughout this work.

\paragraph{Users}
Based on perception studies, users place higher significance on privacy than developers~\cite{senarath2018understanding,hatamian2019revealing} and are largely concerned about the privacy of long-term identifying information such as email addresses and phone numbers, their browsing content and patterns, and about privacy from voice-recognizing personal assistants~\cite{van2018security,lau2018alexa}.

\medskip \noindent
\textbf{Who is responsible for security and privacy?}
While familiarity with technology correlates with use of proactive privacy safeguards~\cite{van2014studying}, not all users have such expertise, and some have even received recommendations conflicting with best practices~\cite{privacy_advice}. The burden of privacy should not fall most heavily on the users of devices. Protection \textit{by default} avoids the limitations of user interventions~\cite{zurko1996user,adams1999users,whitten1999johnny} or providing security advice~\cite{privacy_advice}, but can be complex or costly. Default protections also ameliorate the scrutiny opt-in measures may attract~\cite{xkeyscore_tor}. The clear takeaway from usable security research is that adapting technical measures, rather than requiring users to adapt, improves privacy outcomes overall.

\subsection{Known Extraction Techniques}
In this work, we consider state-of-the-art confidentiality mechanisms in the literature and in industry. To contextualize these mechanisms, we provide an abridged historical summary of known data extraction techniques. We refer the reader to~\cite{full_paper} for an exhaustive history.

Almost as soon as smartphones were introduced and popularized, methods for extracting data from them arose~\cite{full_paper}. Early methods such as ``chip-off'' extraction of storage media pushed providers to deploy storage encryption~\cite{fukami2017improving}, and later to cryptographically enforce user authentication by requiring the passcode to derive decryption keys~\cite{apple_passcodes_touchid_faceid,aosp_filebased}. As attacks evolved to include exploitation of software and hardware subsystems, providers reacted by including encryption of files at runtime, such as in Data Protection on iOS~\cite{apple_platform_security}, among other mitigations.

\paragraph{Physical access} Data extraction from devices is a constantly-evolving practice. Hardware-based extraction methods have continued to evolve in response to provider changes, synthesizing with software compromise to bypass hardware protections before launching an extraction exploit. For example, analysis indicates~\cite{full_paper,graykey_fcc_news,reed2018graykey} that GrayKey, a proprietary tool made by US company GrayShift~\cite{graykey_fcc}, compromises device subsystems via the USB Lightning port and/or the wireless interfaces on iPhones to launch a passcode-guessing attack~\cite{graykey_fcc,graykey_fcc_news}. In general, these techniques require significant investment to develop, but then may be re-used indefinitely, massively impacting users' data confidentiality.

\paragraph{Cloud access} Cloud extraction methods have arisen as a result of increasing data storage on remote servers. Whether by password or server compromise through software exploits, insider threats, or legal requests, unencrypted data on provider servers, and encrypted data where keys are not user-controlled, are susceptible to attack, and indeed compromised in practice using these techniques~\cite{pi_cloud_extraction,cellebrite_cloud,mennicloud2020,icloudleaks,matzner2014privacy}.

\section{Systematization Methodology}\label{sec:methods}

\paragraph{Organization} We divide our analysis between on-device (\S\ref{sec:ondevice}) and cloud-based (\S\ref{sec:offdevice}) approaches, despite thematically similar challenges, due to their significant variation in functional requirements. Few technical mechanisms in the literature holistically approach \textit{\threatmodelname} either on-device or in the cloud; we therefore develop our taxonomy using interrelated but distinct problem areas. We then evaluate mechanisms via concrete criteria, referencing the directly related adversary capabilities, and highlighting limitations which motivate future work. For each section of our systematization we provide a table visually summarizing the findings of our analysis, and we close each section with motivation and directions for future research.

\paragraph{Criteria for evaluation} Our criteria analyze the extent to which data confidentiality approaches satisfy the privacy requirements of the post-compromise threat model, which are derived directly from the adversarial capabilities discussed in~\S\ref{sec:threatmodels}.

When using encryption for data confidentiality, two defining characteristics arise: what data is encrypted at a given time, and how are encryption keys stored or derived. As a result, we naturally organize the literature into encryption and authentication solutions. In the cloud setting, an additional requirement arises: transparency, that some claimed behavior, cryptographic key, or identity is correct. Within these categories, we evaluate state-of-the-art approaches in the literature and in industry solutions. 

Each confidentiality mechanism may impede or even be mutually exclusive with some functionality; such trade-offs are indeed common~\cite{security_v_functionality}.
Following~\cite{e2e_sok}, we extend our analysis to include usability (for users) and adoptability (for providers). For users, these factors intuitively include performance and ease of use~\cite{zurko1996user,adams1999users,whitten1999johnny}. For providers, they include costs of implementation, and for cloud services the costs of maintenance~\cite{green2016developers}.

\paragraph{Informed evaluation} This work draws from both theory and practice to analyze the research literature. Our systematization is informed by consideration of {\em real, deployed} systems, public record documents, news stories, blog posts, and official device manufacturer and cloud provider documentation. Further, we leverage the recent effort of Zinkus et al.~\cite{full_paper}, which provides an extensive collection of analyses of industry and practice in mobile device data confidentiality.

\section{Threat Model}\label{sec:threatmodels}

\begin{table}[t]
    \centering
    \caption{Classes of \textit{Data Extraction Adversary} Capabilities}\label{tab:capabilities}
    \begin{tabular}{l||l}
        \textbf{Physical (Device) Access} (\S\ref{cap:phys}) & \textbf{Cloud Access} (\S\ref{cap:cloud}) \\\hline
        Logical extraction & Cloud extraction \\
        Manual extraction & Compelled data retrieval \\
        Physical media access & Selective compromise \\
        Compelled decryption & Compelled omissions \\
    \end{tabular}
\end{table}

Due to system complexity and requirements, software and hardware security remain fundamentally difficult, with extensive vulnerabilities reported in each iteration of devices~\cite{full_paper,apple_security_updates} and in their wired~\cite{lu2019salaxy,pangu_bh,checkm8,aleph_qcomm_edl_1} and 
wireless~\cite{artenstein2017broadpwn,xu2019badbluetooth} peripherals. Data extraction tools rely on these vulnerabilities~\cite{full_paper,elcomsoft_methods,reed2018graykey} and have remained successful in practice~\cite{full_paper,upturn_mass_extraction,vice_db,pi_deepdive}, therefore our analysis focuses squarely on \textit{what confidentiality remains} when these protections are bypassed.

\paragraph{Defining the model} We propose a novel threat model, the \textit{data extraction adversary}, which captures the practical realities of data extraction and post-compromise confidentiality. Prior models in the literature fall short of the full capabilities of such an adversary, either lacking consideration of the cloud~\cite{vidas2011all}, of physical access and software compromise~\cite{dolev1983security,bellare1993entity}, or of the subtlety of trusted providers who may be attacked or subpoenaed~\cite{goldreich2009foundations}. Therefore, we enumerate the following real-world capabilities to elucidate this emerging threat model faced by mobile devices. Table~\ref{tab:capabilities} provides a summary which we expand upon in~\S\ref{cap:phys} and~\S\ref{cap:cloud} to describe ten capabilities which guide our systematization. We strongly urge consideration of these capabilities in future work.

\subsection{The \textit{Data Extraction Adversary}}

\subsubsection{Physical Access}\label{cap:phys} Physical device access can enable physical or logical extraction~\cite{full_paper}. Physical compromise entails analysis of storage hardware, whereas logical compromise relies on exploiting device software, co-opting it to extract data. Passcode guessing attacks can even be launched via server ``farms''~\cite{full_paper}. This extensive access can undermine critical security assumptions of many confidentiality mechanisms, rendering them ineffective. In this work, we assume adversarial capabilities are eventually limited by budget -- that is, we assume some costly attacks such as fully decapsulating (dissolving protective layers with acid) and exploiting running hardware is prohibitively expensive, even for such powerful adversaries. This caveat allows us to consider a wider array of technical approaches which are effective in practice. Data extraction adversaries therefore have the following capabilities relating to physical device access:
\begin{enumerate}[label=\textbf{\arabic*}.]
    \item Logical extraction via device compromise~\cite{pi_deepdive,cellebrite_ufed,elcomsoft_methods}
    \item Manual extraction via passcode compromise~\cite{reed2018graykey,ios_user_passcodes}
    \item Manual extraction via interrogation~\cite{vice_db}
    \item Physical media extraction~\cite{skorobogatov2016bumpy,fukami2017improving}
    \item Compelled decryption via court order~\cite{scheffler2021protecting}
\end{enumerate}

\subsubsection{Cloud Access}\label{cap:cloud}

\paragraph{Unique challenges for cloud data} Cloud data can be broadly categorized into three classes, each with unique requirements: cloud services which compute over user data (potentially in aggregate) and provide functionality to users (e.g. navigation, translation, or search); data synchronization, where devices share live data by communicating with cloud storage; and backup, where device data is stored long-term for potential future recovery. Services which compute over sensitive data naturally risk exfiltration or compromise. Data synchronization requires careful key management for shared, sensitive data. Backups require recoverability even if a device (and any encryption secrets it stored) are lost. We discuss these challenges and the extent to which existing approaches address them in~\S\ref{sec:offdevice}.

\paragraph{Providers as targets} Cloud access via compromised or subpoenaed providers or via device or credential compromise represents a powerful capability. Mobile devices are increasingly integrated with cloud functionality for services like messaging, backup, sharing, and storage~\cite{full_paper,apple_icloud_backup,google_gms}. As a result,
access to cloud services and accounts can expose troves of sensitive data~\cite{matzner2014privacy,icloudleaks,cellebrite_cloud,icloud_leak_hack}. Subpoenas may target specific accounts~\cite{apple_transparency,google_transp_2019}, or even physical locations for a period of time -- including anyone who entered the location during the specified period -- via controversial ``geofence'' warrants~\cite{verge_geofence}. A natural response to these threats would be to simply disable cloud functionality on-device; unfortunately, this is a decreasingly viable solution, as the usability impacts may extend beyond users' expectations~\cite{full_paper}. Data extraction adversaries have the following data compromise capabilities relating to cloud services, storage, and backup:
\begin{enumerate}[label=\textbf{\arabic*}.]
    \item Cloud extraction via device compromise~\cite{pi_cloud_extraction,cellebrite_cloud,elcomsoft_tokens}
    \item Compelled data retrieval via court order~\cite{apple_legal,google_transp_2019,scheffler2021protecting}
    \item Cloud extraction via password compromise~\cite{elcomsoft_tokens,cellebrite_cloud}
\end{enumerate}

\paragraph{Providers as accomplices} Whether considering a provider adversarial or simply in compliance legal requests, one must acknowledge providers' ability to materially modify or omit technical safeguards. Our analysis stops short of allowing arbitrary malicious behavior from providers, as their unique control over core device and cloud functionality leaves little room for tenable mitigations. We therefore add the following capabilities relating to data extraction adversaries and provider access to cloud infrastructure:
\begin{enumerate}[label=\textbf{\arabic*}.]
    \setcounter{enumi}{3}
    \item Selectively modifying protections for individual or groups of users~\cite{apple_chinese_govt}
    \item Incentivizing or requiring providers to omit protections~\cite{mennicloud2020,earn_it}
\end{enumerate}

\section{On-Device \textit{\ThreatModelName}}\label{sec:ondevice}

\paragraph{Problem areas} Despite decades of research and industrial advancements, creating, deploying, and maintaining mobile devices which are simultaneously vulnerability-free and enjoy sufficient (i.e. massively marketable) performance and functionality remains out of reach in practice. For example, despite Apple's stated commitment to security and privacy~\cite{apple_privacy} and nearly peerless financial resources, iOS still regularly admits vulnerabilities up to and including jailbreaks~\cite{checkm8,pangu_news,unc0ver}. This is not a criticism of Apple, but rather evidence of the fundamental difficulty of securing complex systems.

Under our threat model, strong \textit{storage encryption} with keys unavailable to potentially compromised software is the primary means of maintaining confidentiality. Encryption is deeply interrelated with authentication: as encrypted data must be accessible for correct functionality, cryptographic \textit{user authentication} is required to mediate access to encryption keys. We analyze the research literature and real-world systems within these two problem areas to characterize the extent of protection available against data extraction adversaries, highlight limitations in this protection, and motivate and guide future work.

\newcommand{\low}{$\square$}
\newcommand{\med}{$\blackdiamond$}
\newcommand{\high}{$\bigvarstar$}
\bgroup
\def\arraystretch{1.1}
\begin{table*}[!ht]
    \begin{centering}
    \caption{On-Device \textit{\ThreatModelName}}\label{tab:ondevice}
    \begin{tabular}{cccccccccccccccccc}
         & & \multicolumn{2}{c}{\textbf{Implemented}} & \rot{\textbf{Device Comp.}} & \rot{\textbf{Passcode Comp.}} & \rot{\textbf{Interrogation}} & \rot{\textbf{Physical Media}} & \rot{\textbf{Court Order}} & & \rot{\textbf{Performance}} & \rot{\textbf{Ease of Use}} & & & \rot{\textbf{Impl. Cost}} & 
         & & \\\hline
        \textit{Problem Area} & \textit{System} & \textit{iOS} & \textit{Android} & \multicolumn{5}{|c}{\textit{Adversary Capabilities}} & & \multicolumn{3}{|c|}{\textit{Usability}} & & \multicolumn{3}{c}{\textit{Adoptability}}\\\hline
  
        \multirow{16}{*}{\textbf{Storage Enc.}}
        & \textbf{Full-Disk Encryption}~\S\ref{sec:storage_enc}\\\cline{2-2}
        & \textit{Data Protection}~\cite{apple_platform_security} & \checkmark & - & \emptycirc & \emptycirc & \emptycirc & \fullcirc & \emptycirc & & \low & \low & & & \Ua\\
        
        & \textit{\texttt{dm-crypt}}~\cite{aosp_fulldisk,dm-crypt} & - & \checkmark & \emptycirc & \emptycirc & \emptycirc & \fullcirc & \emptycirc & & \med & \med & & & \ua\\
        \\ 
        & \textbf{Run-Time File Encryption}~\S\ref{sec:storage_enc}\\\cline{2-2}
        & \textit{Data Protection}~\cite{apple_platform_security} & \checkmark & - & \halfcirc & \emptycirc & \emptycirc & \halfcirc & \emptycirc & & \low & \low & & & \Ua\\
        & \textit{File-Based Encryption}~\cite{aosp_filebased} & - & \checkmark & \emptycirc & \emptycirc & \emptycirc & \halfcirc & \emptycirc & & \low & \low & & & \Ua\\
        \\
        & \textbf{Secure Deletion}~\S\ref{sec:storage_enc}\\\cline{2-2}
        & \textit{via Trusted Hardware}~\cite{apple_platform_security,aosp_securedelete} & \checkmark & \checkmark & \dotcirc & \dotcirc & \dotcirc & \dotcirc & \dotcirc & & \low & \high & & & \Ua\\
        & \textit{via Encryption}~\cite{YANG2018612} & \xmark & \xmark & \dotcirc & \dotcirc & \dotcirc & \dotcirc & \dotcirc & & \med & \high & & & \ua\\
        \\
        & \textbf{Deniable Encryption}~\S\ref{sec:storage_enc}\\\cline{2-2}
        & \textit{BurnBox}~\cite{tyagi2018burnbox} & \xmark & \xmark & \fullcirc & \fullcirc & \fullcirc & \fullcirc & \emptycirc & & \low & \high & & & \Ua\\
        & \textit{Filesystem PDE}~\cite{jia2017deftl,chang2018user,chen2020infuse} & \xmark & \xmark & \fullcirc & \fullcirc & \dotcirc & \fullcirc & \fullcirc & & \med & \high & & & \Ua\\\\
        \cline{1-16}
        \multirow{14}{*}{\textbf{User Auth.}}
        & \textbf{Biometrics}~\S\ref{sec:device_auth} & \checkmark & \checkmark \\\cline{2-2}
        & \textit{Fingerprint}~\cite{apple_passcodes_touchid_faceid,aosp_fingerprint} & \checkmark & \checkmark & \emptycirc & \emptycirc & \emptycirc & \fullcirc & \emptycirc & & \low & \med & & & \Ua\\
        & \textit{Facial Recognition}~\cite{apple_faceid_security,aosp_faceauth} & \checkmark & \checkmark & \emptycirc & \emptycirc & \emptycirc & \fullcirc & \emptycirc & & \low & \med & & & \Ua\\
        \\
        & \textbf{Passcodes}~\S\ref{sec:device_auth}\\\cline{2-2}
        & \textit{Numeric Passcodes}~\cite{ios_user_passcodes,aosp_authentication} & \checkmark & \checkmark & \emptycirc & \emptycirc & \emptycirc & \fullcirc & \dotcirc & & \low & \low & & & \Ua\\
        & \textit{Patterns}~\cite{aosp_authentication} & \xmark & \checkmark & \emptycirc & \emptycirc & \emptycirc & \fullcirc & \dotcirc & & \low & \low & & & \Ua\\
    & \textit{Arbitrary Passphrases}~\cite{apple_passcodes_touchid_faceid,aosp_gatekeeper} & \checkmark & \checkmark & \emptycirc & \emptycirc & \emptycirc & \fullcirc & \dotcirc & & \low & \med & & & \Ua\\
        \\
        & \textbf{Trusted Hardware}~\S\ref{sec:device_auth}\\\cline{2-2}
        & \textit{TrustZone}~\cite{arm_trustzone,aosp_trusty,qsee_info} & - & \checkmark & \halfcirc & \emptycirc & \emptycirc & \halfcirc & \emptycirc & & \low & \low & & & \Ua\\
        & \textit{StrongBox Keymaster}~\cite{android_dev_keystore,android_dev_piechange} & - & \checkmark & \halfcirc & \emptycirc & \emptycirc & \fullcirc & \emptycirc & & \low & \low & & & \Ua\\
        & \textit{SEP}~\cite{apple_platform_security,demystifying_sep_2016} & \checkmark & - & \halfcirc & \emptycirc & \emptycirc & \fullcirc & \emptycirc & & \low & \low & & & \Ua\\
        \\
        \hline 
    \end{tabular}
    \end{centering}
    \\\textbf{Implemented}: \checkmark~= by the provider; \xmark~= no 1st-party implementation; - = N/A
    \\\textbf{Adversary Capabilities}: Mitigates... \emptycirc~= never; \halfcirc~= partially; \dotcirc~= conditionally; \fullcirc~= completely
    \\\textbf{Performance}: Impact... \low ~= minimal; \med~= noticeable; \high~= significant
    \\\textbf{Ease of Use}: Requires... \low~= no interaction; \med~= user opt-in; \high~= user intervention or configuration
    \\\textbf{Implementation Cost}: Requires... \ua~= specialized software; \Ua~= specialized hardware \& software
\end{table*}
\egroup

\subsection{Evaluation}

\subsubsection{Storage Encryption}\label{sec:storage_enc}

To protect data on a device from extraction, whether by manual analysis of physical storage media~\cite{skorobogatov2016bumpy,fukami2017improving}, or (more commonly~\cite{full_paper}) by logical extraction using a compromised software component or operating system kernel~\cite{reed2018graykey,elcomsoft_methods,elcomsoft_nojb,cellebrite_ufed,cellebrite_advanced_services}, data must be encrypted with keys inaccessible to the extractor. Cryptography is leveraged to create systems with a variety of properties ranging from data indistinguishability to deniability, as discussed in this section.

\paragraph{Full-disk encryption (\ref{cap:phys}--4)} FDE enables a disk to be transparently encrypted and decrypted, providing protection to data at rest when a device is powered off~\cite{essiv_disk_enc}. Cryptographic cipher configurations have emerged and been standardized~\cite{1619} and implemented~\cite{apple_platform_security,aosp_fulldisk} by providers to address challenges of efficiency and security~\cite{liskov2002tweakable,essiv_disk_enc,martin2010xts}, most notably a lack of additional space in software-only implementations to store ciphertext-adjacent data such as encrypted keys, nonces, or authentication tags, and other failures in hardware-backed FDE~\cite{meijer2019self}. Legal analysis indicates that encryption keys derived from secrets the user remembers cannot be compelled in some jurisdictions~\cite{scheffler2021protecting}.

However, the success of FDE in mobile devices is varied~\cite{full_paper,casey2008impact} due to the design of decrypting all data after device startup (unlike run-time file encryption), and adaptation by adversaries. Two key adaptations are $1)$ not allowing mobile devices to discharge~\cite{elcomsoft_methods,elcomsoft_usb,elcomsoft_usb_bypass} and $2)$ obtaining passcodes/passwords via keyloggers, extortionary actions and court orders, searches, or even imprisonment to compel divulging of passwords~\cite{casey2008impact}. These adaptations essentially negate the protective benefits of FDE, especially considering that modern mobile devices are likely powered-on at almost all times.

From a usability perspective, FDE minimally delays startup time. Cryptographic accelerators~\cite{apple_platform_security,basu2019nist} are a vital optimization to maintain performance and power efficiency for FDE, and efficient alternatives have been explored for devices lacking such hardware~\cite{crowley2018adiantum}. Regarding adoptability, implementations of FDE are widely available, such as the Linux kernel module \texttt{dm-crypt}~\cite{dm-crypt,demir2020optimizing}, and providers use these or implement their own for mobile devices~\cite{apple_platform_security,aosp_fulldisk}, although in some cases leave them disabled by default apparently for performance reasons~\cite{elcomsoft_android_encryption}. Implementation is not necessarily straightforward, with misconfigurations exposing immense amounts of data as recently as 2019~\cite{meijer2019self}, and thus implementation complexity cannot be disregarded.

\paragraph{Run-time file encryption (\ref{cap:phys}--1,4)} RTFE brings many advantages of FDE to the powered-on, running system by enabling on-demand decryption and re-encryption of data~\cite{blaze93cryptographic,halcrow2005ecryptfs,apple_platform_security}. The advantages of maintaining data encryption after startup are most clearly reflected in their effect on data extraction: in many cases, RTFE-encrypted data is often protected from extraction in practice, often leaving adversaries only able to access data already decrypted at the time of device seizure~\cite{full_paper}; however, this protection depends heavily on RTFE configurations for different types of data. As with FDE, RTFE is seemingly also protected from legal compulsion in some jurisdictions.

RTFE has one key consideration regarding usability: after device lock, any data configured to be re-encrypted is no longer available for use. As a result, apps and lock screen interfaces must be adapted to account for inaccessible data. This issue is largely mitigated by the ease of user authentication (discussed later in this section) and by user interface design. Finally, as with FDE, the advantages of RTFE come with substantial performance costs which must be mitigated by specialized hardware. Cryptographic accelerators are key components of RTFE due to the just-in-time decryption design.

\paragraph{Secure deletion (\ref{cap:phys}--1,2,3,4,5)} Secure deletion is critical for resilient confidentiality, as deleted files may contain as much or more information (by nature of having been deleted) than existing files on a device. Secure deletion is generally achieved via encryption, using cryptographic accelerators, with random keys which are expunged~\cite{peters2015defy}; recent work enables this on mobile devices (which commonly use NAND flash storage~\cite{210544} which entails specific challenges) including without system-level privileges~\cite{YANG2018612}. Alternatively, functionality of the underlying hardware can provide secure deletion~\cite{reardon2013sok}. Apple iOS provides a mechanism referred to as ``effaceable storage'' for secure deletion of encryption keys~\cite{apple_platform_security}. This mechanism uses a combination approach: encryption of data plus hardware functionality to securely delete encryption keys. Android allows secure deletion of keys in trusted hardware via ``rollback resistance''~\cite{aosp_securedelete}: by preventing rollback of a key deletion operation, an encryption key is purged.

Secure deletion can mitigate the impact of data extraction, whether by software, physical, or even passcode compromise, albeit with three significant limitations: first, a user must know a search is coming to preempt it; second, deleted data is lost; and third, it is not always clear when data is securely deleted. The second limitation can be addressed via cloud backup, which itself has risk. Recoverable secure deletion has been recently contributed to the literature~\cite{tyagi2018burnbox}, but still suffers from the limitation of preemption, and further, creates recovery data to be stored remotely which is susceptible to compromise. Secure deletion is also a critical component of run-time file encryption, as to re-encrypt data, keys and decrypted copies of data must be irrecoverably evicted from storage. Securely deleted data is additionally protected from court compulsion, although this risks charges of destruction of evidence. However, such considerations are beyond the scope of this technical paper.

Once implemented in software, potentially leveraging specialized encryption hardware, secure deletion operates transparently to the user. However, considering the third mentioned limitation, it is not always clear that data has been deleted, and may simply be hidden from the user interface for later deletion~\cite{full_paper}.

\paragraph{Deniable encryption (\ref{cap:phys}--1,2,3,4,5)} Plausibly deniable encryption (PDE) is a line of research which examines how encrypted storage systems can be made \textit{deniable}: storage which can feasibly be hidden from an adversary performing a search which potentially includes device access and even compelled decryptions. Approaches to PDE generally involve metadata hiding~\cite{anderson1998steganographic}: preventing filesystem metadata from betraying the location or existence of hidden data using encryption or apparently-unused storage areas.

PDE can provide a large degree of protection against all kinds of device compromise and search, although this protection is limited in that a user may be required to successfully deceive an adversary, fabricating that no further data exists. As noted by~\cite{tyagi2018burnbox}, this can be a significant limitation, especially for users under duress. Such deniability can even extend to legal compulsion, noting risks of perjury. PDE schemes can even allow ``dummy'' passcode disclosure to revel a subset of protected data to satisfy an authority~\cite{peters2015defy}.

Although neither of the major platforms provide PDE, various approaches have been proposed in the literature. Recent work has brought deniable encryption to NAND flash storage systems~\cite{jia2017deftl}, generally challenging due to wear-leveling systems~\cite{210544}. Since then, contributions have provided ``user-friendly'' switching between deniable and regular encrypted storage~\cite{chang2018user} and methods for plausible deniability of the very presence of a PDE system on a device~\cite{chen2020infuse}. However, even with cryptographic hardware, these systems still burden users with noticeable overhead.

\paragraph{Trusted hardware (\ref{cap:phys}--1,4)} Trusted hardware enables storage and usage of secrets such as encryption keys without leaking them to potentially compromised device software and hardware. Particularly, secrets derived from user authentication (discussed later in this section) are protected from the rest of the system, making storage encryption resilient to attack; for this reason we list trusted hardware under user authentication in Table~\ref{tab:ondevice} despite its relevance across categories. Trusted hardware requires embedded micro-controllers protected from a wide range of attacks, from physical reverse-engineering to software attacks launched by a compromised kernel. These protections include physical hardening and tamper detection~\cite{apple_platform_security}, micro-architectural defenses~\cite{gotzfried2017cache,van2018foreshadow}, replay protection~\cite{apple_platform_security,taylor2016security,arm_trustzone}, encrypted RAM~\cite{apple_platform_security,arm_trustzone}, and minimal ``trusted computing bases'' (TCBs)~\cite{klein2009sel4} reducing code size to minimize possibility of vulnerability. Trusted hardware can then be leveraged for storage encryption to handle decryption keys.

Trusted hardware has been implemented in iOS and Android via specialized hardware and software, referred to respectively as the Secure Enclave Processor (SEP) and its operating system SEPOS, and StrongBox on Android~\cite{apple_platform_security,apple_secureelement,android_dev_keystore}. Some older Android devices instead use TrustZone implementations such as Trusty~\cite{arm_trustzone,aosp_trusty}. Trusted hardware has also been developed independently in the literature: Keystone~\cite{10.1145/3342195.3387532} is a realization of trusted hardware relying on end-to-end hardware verification on the open RISC-V platform~\cite{asanovic2014instruction}.

Once deployed, trusted hardware can operate transparently to the user, and with minimal performance impact. For storage encryption, a critical part of this performance comes from hierarchical key management: trusted hardware generally only stores keys at the root of large hierarchies otherwise stored in main (unprotected) memory. However, these hierarchies are encrypted with the root keys, and thus are protected without overburdening the trusted hardware. The protections trusted hardware relies upon generally scale in terms of performance up to their usage in these embedded platforms, but often not beyond: for example, encrypted RAM is used for embedded systems~\cite{henson2013beyond}, but the performance cost this entails has precluded more general use~\cite{5655081,wurstlein2016exzess}. Trusted hardware also has implications for maintaining the integrity of core system components. For example, a device can employ a hardware root of trust to verify the digital signatures of its firmware and bootloader~\cite{apple_platform_security,aosp_verified}. The device's trusted hardware would stop execution if an unsigned bootloader or firmware update is run, preventing a potentially compromised operating system from accessing user data.

\subsubsection{User Authentication}\label{sec:device_auth}

Authentication on-device is the necessarily complement to encryption: when data access is required, secure authentication mechanisms mediate the derivation or release of encryption keys.

\paragraph{Biometrics (\ref{cap:phys}--4)} Biometrics have been studied in the biology, statistics, and criminology literature for decades~\cite{bhattacharyya2009biometric}. However, with the advent of digital biometrics on mobile devices~\cite{apple_passcodes_touchid_faceid,apple_faceid_security,touchidimpact15,aosp_fingerprint,aosp_faceauth}, their use has rapidly transitioned from exceptional to ubiquitous.

Android and iOS both provide biometric authentication in the forms of fingerprint and facial recognition~\cite{apple_platform_security,apple_passcodes_touchid_faceid,aosp_fingerprint,aosp_faceauth}. Apple claims~\cite{apple_passcodes_touchid_faceid} that biometrics encourage users to select stronger passcodes, but this claim has been questioned in the literature~\cite{touchidimpact15}. Recent advances in biometrics have enabled authentication via hand geometry~\cite{barra2019hand}, palm print~\cite{ungureanu2017unconstrained}, iris~\cite{rattani2017online,galdi2017fire,abate2017kurtosis}, periocular~\cite{ahuja2017convolutional,alonso2017log}, and even electrocardiographic identification~\cite{tan2017toward}. Despite this diversity, biometrics all rely on the body and so their threat models and achieved security vary only slightly.

Despite inherent (sensing) hardware complexity and the challenge of securely storing fingerprint and facial recognition data~\cite{apple_faceid_security}, modern devices have successfully deployed trusted hardware to implement biometrics which has resulted in more seamless authentication. In combination with trusted hardware~\cite{apple_platform_security,aosp_gatekeeper,aosp_fingerprint}, biometrics enable protection against logical and physical extraction attacks which lack access to the user. An important caveat to biometric authentication is that while passcodes and other memorized secrets generally cannot be legally compelled in some jurisdictions (e.g., the Fifth Amendment in the US~\cite{eff_5th,scheffler2021protecting}), biometrics generally can be, incurring notable risk.

\paragraph{Passcodes and variants (\ref{cap:phys}--4,5)} Passcodes and similar systems have become the norm for mobile device authentication~\cite{apple_platform_security,aosp_authentication}. When combined with the secure processing, attempt-limiting, and time-delaying functionalities of trusted hardware, passcodes become a convenient and secure mechanism from which encryption keys can be derived to protect data. Further, passcodes and other memorized secrets are often protected from legal compulsion via court order due to the Fifth Amendment or similar laws~\cite{eff_5th,scheffler2021protecting}.

iOS and Android support authentication in the form of PINs, pattern-based codes, and arbitrary passcodes~\cite{apple_platform_security,aosp_authentication}. Short passcodes such as historically default 4-digit passcodes on iOS~\cite{apple_security_guides} and even 6-digit modern defaults~\cite{apple_platform_security} and patterns on Android offer limited protection against brute-force attacks~\cite{aviv2015bigger,ios_poweroff,skorobogatov2016bumpy,reed2018graykey}, common passcode guessing~\cite{ios_user_passcodes}, or other similar attacks~\cite{aviv2010smudge,patternlistener_patterns}. Indeed, when trusted hardware fails to enforce guessing limits and delays, long passcodes are a last line of defense~\cite{reed2018graykey,vice_db}. Unfortunately, choosing weak passcodes is a common user behavior~\cite{ios_user_passcodes,pattern_password_security}.

\paragraph{Trusted hardware (\ref{cap:phys}--1,4)} Specifically for user authentication, trusted hardware can be leveraged to safely handle user authentication data such as a passcode or biometric measurement. Trusted hardware can also facilitate time-based delays (such as using PBKDF2~\cite{pbkdf2}), enforce guessing limits, and even erase data when brute-forcing is detected~\cite{full_paper,apple_platform_security}. This adds an additional challenge for data extraction, requiring attackers to bypass trusted hardware~\cite{reed2018graykey} to perform password guessing. The high performance of these mechanisms is derived from direct integration with biometric measurement hardware~\cite{apple_platform_security,aosp_gatekeeper,aosp_fingerprint}. Refer to \S\ref{sec:storage_enc} for general evaluation of trusted hardware in the on-device setting, which we omit here for brevity.

\subsection{Analysis}

Extensive engineering and research effort has been undertaken to secure mobile devices. However, there are practical limitations to the \textit{\threatmodelname} of modern mobile devices, and these limitations highlight potential directions for future research.

\paragraph{Under-utilized run-time file encryption} Although RTFE is a powerful mitigation for \textit{\threatmodelname}, it is significantly hampered in practice by insecure defaults. iOS Data Protection~\cite{apple_platform_security} and Android file-based encryption~\cite{aosp_filebased} defaults to decrypting data after the \textit{initial} user authentication since startup, or ``after first unlock'' (AFU). In practice, data protection classes on iOS are integral to whether or not data is successfully extracted via logical compromise~\cite{full_paper}. By placing Data Protection classes in the hands of developers, some apps can opt-in to protect their users, but lacking secure defaults this mitigation does not apply to vast amounts of user data~\cite{apple_platform_security}. App instrumentation and file access tracking could dramatically improve RTFE by automatically opting unused data into more encrypted Data Protection classes without impacting performance.

\paragraph{Passcodes as single points of failure} If a device is unlocked, or worse, if the passcode is known, device access is nearly unbounded~\cite{elcomsoft_knownpasscode}. There is clearly much work still to be done in understanding user decisions and encouraging stronger passcodes, but the burden should not lie solely on users. Biometrics have had a massive impact on the way people access their devices, but seem not to have improved underlying passcode strength and currently face legal compulsion in the US~\cite{ios_user_passcodes}.

Alternative authentication schemes can provide protection for data even after passcode compromise. These schemes rely on relocating secrets to other parties/devices which can be retrieved for authentication. For example, secret sharing schemes~\cite{shamir79how} allow a secret to be divided into individually useless parts, and recombined efficiently. However, the twin problems of distributing shares and the frequency of authentication render such techniques impractical. Users can opt to relocate secrets to secure yet accessible external devices instead, such as cryptographic hardware tokens (also called 2FA tokens, for two-factor authentication)~\cite{delaune2008formal}. This approach can protect data in case of passcode compromise, but confers costs of inconvenience due to additional overhead, and potential data loss if the hardware device is lost or fails without a backup. Presently, mobile platforms lack support for 2FA in device unlock (despite supporting it in other cases, e.g. web or app authentication~\cite{android_yubikey}). Any such implementation would likely rely on trusted hardware to perform authentication given the user passcode/biometric in addition to the second factor.

\paragraph{Vulnerabilities in trusted hardware} Trusted hardware is heavily relied upon for \textit{\threatmodelname}. There is evidence that trusted hardware on mobile devices has been exploited in practice~\cite{full_paper,reed2018graykey} to enable passcode brute-force attacks. Additionally, Apple recently updated the trusted hardware components which mitigate replay attacks~\cite{apple_platform_security}, and thus we can infer that previous brute-force exploits~\cite{reed2018graykey} may have exploited this subsystem. The SEP API has been fully reverse-engineered~\cite{demystifying_sep_2016}, and therefore may be an appropriate target for formal specification and verification, in an effort to provably mitigate such vulnerabilities in the future.

\paragraph{Formal lower-bound analysis of deniable encryption} The PDE literature includes extensive adversarial formalizations and security analysis. However, future work in lower-bounds proofs may guide PDE design towards optimality. Following the insights of Tyagi et al.~\cite{tyagi2018burnbox}, analysis of oblivious RAM (ORAM)~\cite{oram_lb} may offer promising analogues to aid this work.

\paragraph{Continuous authentication} Continuous authentication systems synthesize biometric and other sensor measurements to confirm that the authenticated user is still operating the mobile device. Google Smart Lock~\cite{gcloud_suite} uses location, proximity detection through motion, reachability of trusted network devices, and voice recognition to prevent locking the device when the user is continuously identified. User behavior such as touchscreen interaction, walking gait, and app usage can be similarly used for authentication on mobile devices~\cite{MAHFOUZ201728}. Future work could apply these mechanisms to run-time file encryption or to authenticate sensitive user actions.

\paragraph{Cryptographic warrant enforcement} Law enforcement search of mobile devices is commonly governed by warrants which approve the extent of access allowed. There are no extant cryptographic methods for efficiently enforcing such a policy, but achieving such a system is an open problem of great promise. Witness encryption~\cite{garg2013witness} could be leveraged to decrypt only data which falls under a set of predicates determined by a warrant. If this or another cryptographic system could be efficiently realized, the need for law enforcement transparency could be significantly reduced. Initial work in the literature explores this possibility~\cite{arleas}.

\section{Cloud \textit{\ThreatModelName}}\label{sec:offdevice}

\paragraph{Problem Areas} We analyze cloud protection mechanisms from the research literature and from industry which protect data targeted by extraction adversaries (\S\ref{sec:data}) by addressing extraction capabilities (\S\ref{cap:cloud}). \textit{User-controlled encryption} enables sensitive data to remain protected, as cloud providers lack access to encryption keys. To maintain functionality, data must be decrypt-able, and thus requires mediated access via cryptographic \textit{user-to-cloud authentication}. Finally, cloud services may be used to execute sensitive functionality. \textit{Transparency} enables a user/device to ensure correct execution or minimize the impact of malicious parties. Exploiting device trust in providers, a data extraction attack could e.g. replace an HSM IP address with that of an unprotected server; transparency in the setting of \textit{\threatmodelname} replaces trust in providers with verifiability, specifically of cryptographic keys and identities which are required to bootstrap further security. In the evaluation which follows, we analyze the research literature and real-world implemented systems within these three problem areas to characterize the extent of protection available against data extraction adversaries, highlight limitations, and motivate and guide future work.

\bgroup
\def\arraystretch{1.1}
\begin{table*}[ht]
    \begin{centering}
    \caption{Cloud \textit{\ThreatModelName}}\label{tab:offdevice}
    \begin{tabular}{cccccccccccccccccc}
         & & \multicolumn{2}{c}{\textbf{Implemented}} & \rot{\textbf{Device Comp.}} & \rot{\textbf{Court Order}} & \rot{\textbf{Password Comp.}} & \rot{\textbf{Selective Mod.}} & \rot{\textbf{Omission}} & & \rot{\textbf{Performance}} & \rot{\textbf{Ease of Use}} & & & \rot{\textbf{Impl. Cost}} & \rot{\textbf{Maint. Cost}} & & \\\hline
        \textit{Problem Area} & \textit{System} & \textit{Apple} & \textit{Google} & \multicolumn{5}{|c}{\textit{Adversary Capabilities}} & & \multicolumn{3}{|c|}{\textit{Usability}} & & \multicolumn{3}{c}{\textit{Adoptability}}\\\hline
        
        \multirow{7}{*}{\textbf{User-Controlled Enc.}}
        & \textbf{Key Agreement}~\S\ref{sec:cloud_uce}\\\cline{2-2}
        & \textit{Apple Handoff}~\cite{apple_handoff} & \checkmark & - & \emptycirc & \fullcirc & \dotcirc & \emptycirc & \emptycirc & & \low & \low & & & \ua & \da \\\\

        & \textbf{Trusted Hardware}~\S\ref{sec:cloud_uce}\\\cline{2-2}
        & \textit{Titan}~\cite{titan_in_depth} & - & \checkmark & \halfcirc & \fullcirc & \emptycirc & \halfcirc & \halfcirc & & \low & \low & & & \Ua & \ua \\
        & \textit{iCloud Keychain}~\cite{apple_icloud_security} & \checkmark & - & \halfcirc & \fullcirc & \emptycirc & \halfcirc & \halfcirc & & \low & \low & & & \Ua & \ua \\
        \\
        \cline{1-16}
        
        \multirow{9}{*}{\textbf{User-to-Cloud Auth.}}
        & \textbf{PAKE}~\S\ref{sec:cloud_utca}\\\cline{2-2}
        & \textit{OPAQUE}~\cite{jarecki2018opaque}  & \xmark & \xmark & \emptycirc & \fullcirc & \emptycirc & \halfcirc & \halfcirc & & \low & \low & & & \ua & \da \\
        & \textit{SRP}~\cite{apple_platform_security,SRP} & \checkmark & \xmark & \emptycirc & \fullcirc & \emptycirc & \halfcirc & \halfcirc & & \low & \low & & & \ua & \da \\
        \\
        & \textbf{2FA}~\S\ref{sec:cloud_utca}\\\cline{2-2}
        & \textit{OTP}~\cite{m2011totp,m2005hotp} & \xmark & \checkmark & \fullcirc & \emptycirc & \fullcirc & \emptycirc & \emptycirc & & \low & \med & & & \da & \da \\
        & \textit{PKCS \#11}~\cite{delaune2008formal} & \xmark & \xmark & \fullcirc & \dotcirc & \fullcirc & \emptycirc & \emptycirc & & \low & \high & & & \ua & \Da \\
        & \textit{Apple 2FA}~\cite{apple_tfa} & \checkmark & - & \fullcirc & \dotcirc & \fullcirc & \emptycirc & \emptycirc & & \low & \med & & & \ua & \da \\
        \\
        \cline{1-16}
        
        \multirow{7}{*}{\textbf{Transparency}}
        & \textbf{Decentralization}~\S\ref{sec:transparency}\\\cline{2-2}
        & \textit{Blockchain}~\cite{benet2018filecoin} & \xmark & \xmark & \halfcirc & \halfcirc & \emptycirc & \fullcirc & \fullcirc & & \high & \high & & & \ua & \Uda \\
        & \textit{P2P Networks}~\cite{benet2014ipfs} & \xmark & \xmark & \halfcirc & \halfcirc & \emptycirc & \fullcirc & \fullcirc & & \high & \high & & & \ua & \uda \\
        \\
        & \textbf{Transparency Logging}~\S\ref{sec:transparency}\\\cline{2-2}
        & \textit{Trillian}~\cite{trillian} & \xmark & \xmark & \halfcirc & - & \emptycirc & \halfcirc & \fullcirc & & \med & \high & & & \ua & \ua \\
        \\        
        \hline
    \end{tabular}
    \end{centering}
    \\\textbf{Implemented}: \checkmark~= by the provider; \xmark~= no 1st-party implementation; - = N/A
    \\\textbf{Adversary Capabilities}: Mitigates... \emptycirc~= never; \halfcirc~= partially; \dotcirc~= conditionally; \fullcirc~= completely
    \\\textbf{Performance}: Impact... \low~= minimal; \med~= noticeable; \high~= significant
    \\\textbf{Ease of Use}: Requires... \low~= no interaction; \med~= user opt-in; \high~= user intervention or configuration
    \\\textbf{Implementation Cost}: Requires... \ua~= specialized software; \Ua~= specialized hardware \& software; \da~= neither
    \\\textbf{Maintenance Cost}: \Da~= negligible; \da~= low; \ua~= high; \uda~= variable; \Uda~= highly variable
\end{table*}
\egroup

\subsection{Evaluation}

\subsubsection{User-Controlled Encryption for Cloud Data}\label{sec:cloud_uce}

Encryption technologies, such as symmetric ciphers~\cite{aes}, are well-understood and robustly implemented. As a result, the problem of user-controlled encryption in the cloud reduces to one of encryption key management. Provider servers may be compromised by data extraction adversaries, thus encryption keys must be kept exclusively on-device or within trusted hardware. For cloud services which compute over user data, risk must either simply be accepted or mitigated through complex techniques such as homomorphic encryption~\cite{gentry2009fully}. For data synchronization between user devices or backup services, approaches discussed in this section offer trade-offs between confidentiality and complexity, ultimately providing at best partial protection from extraction attacks.

\paragraph{Key agreement (\ref{cap:cloud}--2,3)} For scenarios where cloud data is stored only for purposes of synchronizing multiple live devices, it is possible to derive high-entropy keys between devices using cryptographic key agreement techniques, such as those already used for end-to-end encrypted messaging systems. Indeed, Apple already supports the derivation of pairwise keys as part of its Handoff service~\cite{apple_platform_security}, which is a system that synchronizes data between devices via Bluetooth and WiFi. The advantage of using this approach to deriving keys for cloud services is that the cloud provider never learns the shared device keys. This assumes that the provider does not selectively modify or omit the key agreement protocol, e.g. by tampering with the identity and key distribution services. See \S\ref{sec:transparency} for more on preventing such attacks. The disadvantage of this approach is that it is useful primarily for {synchronizing} data between functioning devices: this approach does not support device backup in the event that all user devices (and hence keys) become unavailable.

\paragraph{Trusted hardware (\ref{cap:cloud}--1,2,4,5)} Trusted cloud hardware allows users to compute privately even when they do not fully trust a cloud provider~\cite{smith1998trusting,irvine2007trusted}. In practice, trusted cloud hardware takes the form of hardware security modules (HSMs) deployed by providers, which generally execute pre-determined functionalities such as encryption or password verification~\cite{apple_platform_security,apple_bh_2016,titan_in_depth,google_titanm}. Trusted hardware enables user-controlled encryption of cloud-stored data without requiring users to memorize or store high-entropy secrets, either through \textit{password strengthening}~\cite{manber1996simple,abadi1997strengthening,kelsey1997secure,halderman2005convenient}, cryptographically mixing the user password (or a derivation thereof) with high-entropy secrets resulting in a secure encryption key unknown to the cloud provider, or via password-authenticated key exchange (\S\ref{sec:cloud_utca}). A key feature of these devices is that they can enforce {\em guessing limits} to prevent dictionary attacks. This approach enables \textit{user-controlled} encryption while mitigating the need for users to memorize or store high-entropy secrets, which may be an untenable UX requirement~\cite{zurko1996user}.

In Apple iCloud, trusted hardware is also used to manage a list of ``trusted devices'' per-account, a mechanism which can be used to share or revoke access to encryption keys among user devices without disclosing them to Apple~\cite{apple_platform_security,apple_icloud_security}. Further, HSMs can cryptographically authenticate their code. Therefore, if the user trusts that Apple correctly implements iCloud Keychain functionality~\cite{apple_bh_2016}, they can rely on this authentication to prevent modifications to that functionality. However, the user has no way to verify they are communicating with an Apple HSM, and thus must implicitly rely on the provider. This caveat has particular relevance in light of Apple's agreement with the Chinese government to move iCloud encryption keys to Chinese servers~\cite{apple_chinese_govt}. For Google Mobile Services (GMS), the Titan HSM system~\cite{titan_in_depth,google_titanm} is used to protect some mobile device backups~\cite{kensinger2018google} using the user's device authentication credential in a similar entropy-stretching design.

The use of trusted cloud hardware poses many challenges for providers. These include the need to prove to users that the hardware is correctly implemented and cannot be re-programmed, as well as many additional deployment challenges around replication and availability (see \S\ref{sec:cloud_discussion} for further discussion).

\subsubsection{User-to-Cloud Authentication}\label{sec:cloud_utca}

\paragraph{Password-authenticated key exchange (\ref{cap:cloud}--2,4,5)} PAKE is a cryptographic protocol in which communicating parties with a shared, low-entropy secret (a password) are able to securely derive a shared high-entropy secret key. In the cloud setting, asymmetric PAKE (aPAKE) is of particular relevance: the user authenticates with their password, which has been previously registered with the cloud (ideally in trusted hardware, per~\S\ref{sec:cloud_uce}) without revealing it, while the cloud is authenticated either implicitly or through public-key infrastructure. Recent cryptographic results in (a)PAKE have achieved seemingly optimal communication complexity~\cite{mcquoid2020minimal}, and even pre-computation attack resistance in the OPAQUE protocol~\cite{jarecki2018opaque}.

Apple has implemented one such aPAKE in iCloud~\cite{apple_platform_security}. iOS users execute the Secure Remote Password protocol (SRP)~\cite{SRP} rather than traditional password authentication to authenticate to iCloud HSMs. As compared with OPAQUE, SRP is not pre-computation attack resistant, and lacks a formal proof of security with standard assumptions. The SRP interaction is transparent to users and confers negligible performance differences, but fully hides the user's iCloud password from Apple servers. Implementing SRP required up-front investment from Apple, but no significant additional upkeep. Although in theory users could notice if Apple servers discontinued use of SRP, the technical knowledge required implies that Apple could, if compelled, selectively remove this feature covertly. However, this mechanism partially maintains security against server-side modifications or omissions as it does not leak the user password.

\paragraph{Two-factor authentication (\ref{cap:cloud}--1,2,3)} 2FA for cloud services can maintain security in the event of password or device compromise. 2FA via cryptographic hardware tokens has been standardized, and the research literature contains extensive formalization and analysis of these standards~\cite{delaune2008formal,clulow2003security,bortolozzo2010attacking}.

Google has implemented 2FA in multiple forms, allowing for hardware tokens, and time- and HMAC-based one-time passwords (OTP)~\cite{google_tfa}. These implementations provide users opt-in measures to increase the security of their cloud accounts, however, OTP secrets are likely accessible to providers by design~\cite{m2011totp,m2005hotp}. These mechanisms require little upkeep, and open-source implementations are readily available. Apple's implementation of 2FA stands out, in that it interacts with their trusted device infrastructure~\cite{apple_tfa} and therefore \textit{may} rely on secret stored in user devices rather than in the cloud, however, this functionality is not documented.

\paragraph{Trusted hardware (\ref{cap:cloud}--1,4,5)} Specific to user-to-cloud authentication, cloud HSMs enable secure storage of user registration information. However, novel and compelling challenges emerge when considering questions of scale: multi-HSM consistency over user authentication data, for example to support load balancing amongst HSMs, has only recently received treatment in practice, and has received almost no formal analysis. Load balancing among HSMs requires secure sharing of not only (static) user authentication data but also (likely dynamic) state associated with the user, such as remaining login attempts. Signal, the end-to-end encrypted messaging platform, recently applied distributed consensus~\cite{lamport1982byzantine,ongaro2014search} to achieve HSM consistency for encrypted backups~\cite{signal_svr}. Apple pre-provisions HSM clusters rather than dynamically scaling~\cite{apple_bh_2016}. Refer to~\S\ref{sec:cloud_uce} for general evaluation of trusted cloud hardware which we omit here for brevity.

\subsubsection{Transparency of Keys and Identities}\label{sec:transparency}

\paragraph{Transparency logs (\ref{cap:cloud}--4,5)}
Providers are trusted to deploy HSMs and manage encryption keys on behalf of their users. Therefore, data extraction attacks launched by or with access to provider systems may exploit this trust. For example, a device could be given the address of a malicious server in place of an HSM. Transparency logs are designed to enforce honest provider behavior by requiring validation against a public log. In the HSM example, a device would be able to validate the identity of the server against such a log. This enforcement has limitations, as new log entries can not necessarily be validated in real-time or with high degrees of certainty, but the approach helps users ensure that their view of cloud systems is consistent. These systems do not alone explicitly prevent surreptitious behavior, but combined with device-side verification of public transparency data, transparency logs can mitigate covert attacks including those against key management, software update delivery, and HSM provisioning.

Public ledgers~\cite{kaptchuk20new} can provide resilient, distributed storage for transparency logs, and facilitate enforcement of transparency. This model has seen success in Google's Certificate Transparency (CT)~\cite{laurie14certificate}, a system for auditing the distribution of TLS certificates. Google has since generalized their implementation to support general verifiable data structures~\cite{trillian}. Transparency logs are also used to validate the provenance of keys~\cite{coinks_key_transparency,google_key_transparency} and software packages~\cite{go_transparency}. Transparency log systems require initial implementation investment, but can provide verification by default and can rely on decentralized networks to avoid impacting performance for users or cloud providers.

\paragraph{Decentralization (\ref{cap:cloud}--1,2,4,5)} Another approach to enforcing transparency is to outsource services to decentralized peer networks. With sufficient decentralization, individual server compromise is ineffective, and court orders become infeasible to coordinate and enforce. However, decentralization can incur significant performance overhead, e.g. by collecting shards of user data from nodes potentially across the world. Decentralized networks often require consensus, implying further overhead.

Due to complexity and requiring coordination, decentralization is rarely used to replace cloud services. Apple has implemented a peer-to-peer service for finding lost devices via Bluetooth broadcasts while maintaining a degree of location privacy~\cite{findmy,heinrich2021can}. Current work is largely focused on blockchains, in which append-only public ledgers are stored using consensus protocols~\cite{kaptchuk20new,nakamoto2019bitcoin}. Whether blockchain-based~\cite{benet2018filecoin} or not~\cite{benet2014ipfs}, decentralized data storage systems have potential to mitigate the privacy concerns of provider-controlled cloud servers while relieving providers of maintenance costs.

\paragraph{Trusted hardware (\ref{cap:cloud}--2,4,5)} Specific to transparency, trusted hardware facilitates confidentiality by cryptographically verifying code. Providers leverage this feature to remove their own ability to surreptitiously access data or modify functionality: Apple, for example, claims to have destroyed the code signing keys for iCloud Keychain HSMs~\cite{apple_bh_2016}. However, users generally cannot determine that they are communicating with a correct HSM in practice: thus, this design only provides partial mitigation against many data extraction capabilities. HSMs rely on user authentication to provide data records, and therefore password compromise completely negates their benefit. Refer to~\S\ref{sec:cloud_uce} for general evaluation of trusted cloud hardware which we omit here for brevity.

\subsection{Analysis}\label{sec:cloud_discussion}

Cloud data confidentiality mechanisms center around removing implicit trust in providers. Due to the practical realities of data extraction adversaries, remaining trust creates disconnects in data confidentiality which imply promising directions for future work.

\paragraph{Open challenges for cloud data} Sensitive data in the cloud faces numerous challenges pertaining to confidentiality. Cloud services which compute over user data to provide functionality represent direct risk to confidentiality unless complex approaches such as fully-homomorphic encryption~\cite{gentry2009fully} are adopted by providers. Data synchronization services must authenticate devices and establish encryption keys from low-entropy user credentials via aPAKE or password strengthening, and must safely store these keys (generally, via trusted hardware). Implementing the key agreement model of Firefox Sync~\cite{ff_sync} is a promising engineering solution to mitigate a subset of extraction attacks at relatively low cost. Cloud backups, a prevalent target for subpoena~\cite{upturn_mass_extraction,full_paper}, must also be safely stored while being recoverable even if encryption secrets, devices, and/or passwords are forgotten or lost. Trusted hardware, in combination with other mitigations, has significantly improved these open problems in practice, but creates new challenges: HSM reprovisioning and scaling have been initially explored but lack formal analysis. As a result, maintaining data confidentiality in the cloud in the post-compromise setting remains a promising direction for impactful future work.

\paragraph{Data recoverability and backups} The realities of portable devices and human nature create risk of accidental data loss. As a result, providers offer cloud backup services, encrypting these backups with keys they control~\cite{apple_icloud_backup,android_dev_autobackup}. Ostensibly this is to mitigate loss due to forgotten passcodes, but additional pressure from law enforcement is also allegedly a factor~\cite{mennicloud2020,pi_cloud_extraction}. Recoverability seems to imply a dilemma for cloud backup encryption: backups cannot be simultaneously fully recoverable (including upon passcode loss) yet provider-inaccessible (through user-controlled encryption). Biometric-derived encryption keys may hold promise in resolving this dilemma, and a number of user interface solutions might mitigate its impact. Due to the high frequency of subpoena~\cite{upturn_mass_extraction, pi_cloud_extraction} cryptographic tools such as functional encryption~\cite{boneh2011functional} may also hold promise in improving privacy while enabling lawful search.

\paragraph{Trusted hardware} In practice, trusted hardware plays a vital role in addressing the duality of trust in providers. By deploying trusted hardware, a provider can remove their own ability to circumvent privacy mechanisms. As such, trusted hardware is thoroughly examined in this systematization, and we identify and formalize the \textit{\defenseparadigm} mitigation paradigm as an emerging pattern. Analyzing the extent and limitations of this paradigm is itself an opportunity for future work. Instantiations of this model in practice still implicitly rely on providers: users generally cannot verify HSM instances and cannot distinguish honest reprovisioning from an attack. Closing these gaps represents multiple lines of promising and impactful future work.

\paragraph{Remaining trust in providers} Finally, our threat model calls into question services in which the provider acts as an identity broker, such as with Apple iMessage, Apple FaceTime, and Google Duo, and of security features such as iCloud Keychain. Existing designs leverage the provider to provide efficient distribution of cryptographic material between peers. However, centralized designs, while relatively performant, require trust in the provider. In some cases, security features are intentionally omitted to enable functionality: iCloud Backup~\cite{apple_icloud_backup} and Android Auto-Backup~\cite{android_dev_autobackup} are stored encrypted with keys held \textit{by the providers}, which allows them to restore user data even if a user forgets their passcode. Worse still, trusted hardware attestation keys have leaked~\cite{van2018foreshadow}. Myriad opportunities exist for further reducing unneeded trust in providers: verifying peer identity, validating server behavior or committing it to transparency logs, deploying user-controlled encryption, or decentralizing, even if only among groups of providers. Eliminating this trust will significantly improve resilience of device and cloud systems against even the strongest covert and malicious adversaries.

\section{Conclusion}\label{sec:conclusion}

Modern mobile devices, with their storage, sensing, and connectivity capabilities, are of unparalleled value to adversaries seeking sensitive personal data. Software security techniques in industry and the literature continue to provide complex and comprehensive protections against many attacks, and yet vulnerability-free mobile operating systems remain out of reach. \textit{\Threatmodelname} has therefore risen to paramount importance.

We contribute a novel threat model, informed by the realities of the mobile device ecosystem and formalize an emerging defense paradigm for cloud service. We systematize research, providing a thorough evaluation of the current state of both the research literature and engineering results, as well as motivation and directions toward open questions. The open questions we suggest relate to concretely improving privacy and security through technical measures across various fields of research.

It is our hope that this work facilitates collaboration across academia, industry, and policymaking, and promotes not only research but tangible impact towards \textit{\threatmodelname} for users of mobile devices.

\paragraph{For researchers} In applying cryptographic, security, and privacy research to the mobile setting, researchers should consider the numerous and subtle effects of cloud integration and trusted hardware. As we have enumerated, there are extensive opportunities for novel work across the theory and practice of these emerging settings, and the cloud can provide both significant benefits and unforeseen risks to applied research.

\paragraph{For providers} Users rely on providers for safety and peace of mind. In many regards, providers rise to this mantle by improving security and privacy in devices and services, and contributing to research. In this work we identify a number of gaps between providers' implementations and \textit{\threatmodelname}. Working to address these will create substantial privacy and security benefits for users.

\paragraph{For policymakers} Device and cloud protections, while increasingly robust, are bypassed in practice by lawful and illicit actors alike. Historically, government standards and practices have improved security and privacy, but in recent years we observe a concerning reversal of this trend. A wide range of user data is readily extractable upon legal request. Rather than weakening or banning strong, user-controlled encryption, policymakers should encourage providers to bolster existing defenses and work with partners in academia and industry to find solutions to incorporate lawful access with strong cryptography. 

\section*{Acknowledgments}
 
The authors would like to thank Dr. Mike Rushanan, and Emma Weil and Dr. Harlan Yu of Upturn, for their insightful feedback.

The authors received support from the National Science Foundation under awards CNS-1653110 and CNS-1801479, and from a Google Security \& Privacy Award as well as an ONR award.  Additionally, this material is based upon work supported by DARPA under Contract No. HR001120C0084. Any opinions, findings and conclusions or recommendations expressed in this material are those of the author(s) and do not necessarily reflect the views of the United States Government or DARPA.

\bibliography{main}
\end{document}

%% file: localdefs.tex
\renewcommand{\paragraph}[1]{\medskip \noindent {\bf #1}.}
\long\def\symbolfootnote[#1]#2{\begingroup%
\def\thefootnote{\fnsymbol{footnote}}\footnote[#1]{#2}\endgroup}

\let\latexcite=\cite
\def\cite{\nolinebreak\latexcite}
\let\latexref=\ref
\def\ref{\nolinebreak\latexref}

\usepackage{color}

\newcommand{\ignore}[1]{}

\newcounter{romanlistcounter}
  {\setcounter{romanlistcounter}{0}%
   \begin{list}{\textit{(\roman{romanlistcounter})}}{%
        \usecounter{romanlistcounter}%
      \setlength{\itemsep}{0pc}%
      \setlength{\itemindent}{1pc}%
      \setlength{\topsep}{0pc}%
      \setlength{\mylabelwidth}{3em}}}
  {\end{list}}

\newcounter{alphalistcounter}
  {\setcounter{alphalistcounter}{0}%
   \begin{list}{\textit{(\alph{alphalistcounter})}}{%
        \usecounter{alphalistcounter}%
      \setlength{\itemsep}{0pc}%
      \setlength{\itemindent}{1pc}%
      \setlength{\topsep}{0pc}%
      \setlength{\mylabelwidth}{3em}}}
  {\end{list}}

\def\compactify{\itemsep=0pt \topsep=0pt \partopsep=0pt \parsep=0pt}
\let\latexusecounter=\usecounter

\let\latexusecounter=\usecounter